\documentclass[journal,article,oneauthor,10pt,a4paper]{mdpi_nolinenum} %
\firstpage{1} 
\articlenumber{x}
\doinum{10.3390/------}
\pubvolume{xx}
\pubyear{2016}
\copyrightyear{2016}
\externaleditor{Academic Editor: name}
\history{Received: date; Accepted: date; Published: date}

 \theoremstyle{mdpi}
 \newcounter{thm}
 \newcounter{ex}
 \newcounter{re}
 \theoremstyle{mdpidefinition}

\Title{Spectral and Polarization Signatures of Relativistic Shocks in Blazars}

\Author{Markus B\"ottcher $^{1,2}$}
\AuthorNames{Markus B\"ottcher}

\address{%
$^{1}$ \quad Centre for Space Research, North-West University, Potchefstroom, 2520, South Africa; 
Markus.Bottcher@nwu.ac.za\\
$^{2}$ \quad Department of Physics and Astronomy, Ohio University, Athens, OH 45701, USA
}

\abstract{Relativistic shocks are one of the most plausible sites of the emission of strongly
variable, polarized multi-wavelength emission from relativistic jet sources such as blazars,
via diffusive shock acceleration (DSA) of relativistic particles. This paper summarizes
recent results on a self-consistent coupling of diffusive shock acceleration and radiation
transfer in blazar jets. We demonstrate that the observed spectral energy distributions (SEDs)
of blazars strongly constrain the nature of hydromagnetic turbulence responsible for pitch-angle
scattering by requiring a strongly energy-dependent pitch-angle mean free path. The prominent
soft X-ray excess (``Big Blue Bump'') in the SED of the BL Lac object AO 0235+164 can be
modelled as the signature of bulk Compton scattering of external radiation fields by the 
thermal electron population, which places additional constraints on the level of hydromagnetic
turbulence. It has further been demonstrated that internal shocks propagating in a jet pervaded
by a helical magnetic field naturally produce polarization-angle swings by 180$^o$, in tandem
with multi-wavelength flaring activity, without requiring any helical motion paths or other
asymmetric jet structures. The specific application of this model to 3C279 presents the first
consistent, simultaneous modeling of snap-shot SEDs, multi-wavelength light curves and time-dependent
polarization signatures of a blazar during a polarization-angle (PA) rotation. This model has recently
been generalized to a lepto-hadronic model, in which the high-energy emission is dominated by
proton synchrotron radiation. It is shown that in this case, the high-energy (X-ray and $\gamma$-ray)
polarization signatures are expected to be significantly more stable (not showing PA rotations)
than the low-energy (electron-synchrotron) signatures. 
}


\begin{document}


\section{\label{intro}Introduction}

Ever since the pioneering work of Marscher \& Gear \cite{MG85}, relativistic shocks have been considered one of 
the leading contenders for the location of relativistic particle acceleration, resulting in the observed rapidly 
variable, often highly polarized multi-wavelength emission from blazars. Models focusing on the multi-zone 
radiative transfer problem, given efficient relativistic shock acceleration, have reached an increasing level 
of sophistication \citep[e.g.,][]{Spada01,Sokolov04,Mimica04,SM05,Graff08,BD10,JB11,Chen11,Chen12}, 
resulting in successful fits to multi-wavelength SEDs and light curves of individual flares in blazars. 
In all cited works, however, the details of particle acceleration are not specifically addressed, but
shock acceleration is assumed to result in the injection of relativistic particles, typically with a
power-law distribution in energy. In a recent work, Baring et al. \citep{Baring16} have coupled the
Monte-Carlo simulations of diffusive shock acceleration of Summerlin \& Baring \cite{SB12} with radiative
transfer routines of B\"ottcher et al. \cite{Boettcher13}. Their results are briefly summarized in
Section \ref{DSA}, focusing on implications for the nature of hydromagnetic turbulence facilitating
pitch-angle scattering. 

An important additional observable which is usually neglected in detailed radiative transfer simulations
for blazars, is the linear polarization of synchrotron (and possibly also Compton) emission. A peculiar
feature that has been observed in several cases \citep[e.g.,][]{Marscher08,Abdo10} consists of large
polarization-angle (PA) swings by $\ge 180^o$ asociated with multi-wavelength flaring activity. With 
very few exceptions \citep[e.g.,][]{Marscher14,Zhang14}, theoretical models for synchrotron polarization
variations are purely geometrical, often with an ad-hoc assumption of helical pattern motions guided by 
helical magnetic fields, without consistent considerations of radiation physics or particle dynamics 
\citep[e.g.,][]{Larionov13}. In this context, it should be pointed out that most works modeling the
SEDs of blazars in detail \citep[e.g.,][]{Ghisellini10,Boettcher13} indicate that in the high-energy
emission region, blazar jets are particle dominated (i.e., with sub-equipartition magnetic fields), 
in which case the relativistic plasma is not expected to be guided by the magnetic fields, but 
instead the magnetic fields will follow the plasma motion. Note furthermore that in the case of 
polarization and flux variations dominated by viewing-angle changes, the varibility is expected 
to be dominated by Doppler-factor changes. In this case, the variability is quasi-achromatic, 
with peak flux variations of individual radiation components, $\delta (\nu F_{\nu}^p)$ related 
to peak frequency variations $\delta \nu_p$ as $\delta (\nu F_{\nu}^p) \propto (\delta \nu_p)^4$. 
This contradicts the strongly frequency-dependent flux variability patterns normally observed in 
blazars.

One of the first models that consistently treat radiation physics with time-dependent synchrotron 
polarization signatures, is the Turbulent Extreme Multi-Zone Model (TEMZ) \cite{Marscher14}. Here,
polarization variations are the result of the energization of a small number of radiative cells,
each with randomly oriented (turbulent) magnetic fields. This model has proven very successful in
reproducing stochastic variations of polarization characteristics, and may occasionally even lead
to organized PA swings. In this model, PA swings are not expected to be systematically associated 
with multi-wavelength flaring activity, in agreement with the statistical analysis of a large number
of PA rotations observed by RoboPol \citep{Blinov16}. However, there is also a significant number
of events in which PA rotations do occur in clear correlation with multi-wavelength flares
\citep[e.g.][]{Marscher08,Abdo10}. For this latter case, Zhang et al. \cite{Zhang14,Zhang15}
have developed a model self-consistently tracing particle dynamics, radiation transfer and
resulting polarization characteristics in a shock-in-jet model with an ordered, helical magnetic
field. Their results will be briefly summarized in Section \ref{PAswings}. This model has recently
been generalized to a lepto-hadronic model, which will be presented in Section \ref{leptohad}.

\section{\label{DSA}Diffusive Shock Acceleration in Relativistic Jets}

Diffusive shock acceleration (DSA) is generally considered to be one of the most efficient modes
of particle acceleration at relativistic shocks. Detailed Monte-Carlo (MC) simulations of DSA at
relativistic, oblique shocks have recently been carried out by Summerlin \& Baring \cite{SB12}.
In the DSA scenario, the Fermi-I acceleration process is facilitated by pitch-angle scattering,
causing the guiding centers of particles spiraling along magnetic field lines to reverse their
direction of motion in a stochastic manner. Pitch-angle scattering is parameterized through the
pitch-angle mean-free path $\lambda_{\rm mfp}$ as a (generally energy-dependent) multiple 
$\eta (p)$ of the particle's gyro radius, $r_g = p c / (q B)$, where $p$ is the particle's 
momentum, such that $\lambda_{\rm mfp} = \eta (p) \, r_g$. The energy dependence of the 
mean-free-path parameter $\eta$ is parameterized as a power-law in the particle's momentum,
$\eta (p) = \eta_1 \, p^{\alpha - 1}$, so that $\lambda_{\rm mfp} \propto p^{\alpha}$ and
$\eta_1$ describes the mean free path in the non-relativistic limit, $\gamma \to 1$. 

In a recent paper, Baring et al. \cite{Baring16} have now coupled the MC simulations of DSA to
radiative transport, adopting the routines of \cite{Boettcher13}. To obtain realistic thermal +
non-thermal particle distributions, a high-energy cut-off ($\gamma_{\rm max}$) of the non-thermal 
particle spectra has been evaluated by balancing the acceleration time scale $t_{\rm acc} 
(\gamma_{\rm max}) = \eta (\gamma_{\rm max}) \, t_{\rm gyr} (\gamma_{\rm max})$ with the
radiative energy loss time scale. In the presence of dominant synchrotron losses, this will 
lead to a synchrotron peak energy $E_{\rm sy} \sim 240 \, \delta \, \eta^{-1} (\gamma_{\rm max})$~MeV, 
independent of $B$, where $\delta$ is the Doppler factor. Thus, in order to reproduce a synchrotron 
peak in the IR to soft X-rays, as typical for blazars, $\eta (\gamma_{\rm max})$ has to assume values 
of $\sim 10^4$ -- $10^8$, depending on the location of the synchrotron peak. On the other hand, as 
demonstrated in \cite{SB12}, $\eta_1$ must be significantly smaller than this in order to obtain 
efficient injection of particles out of the thermal pool into the non-thermal acceleration process. 
This indicates that $\eta(\gamma)$ must be strongly energy dependent. 

\begin{figure}[H]
\centering
\includegraphics[width=10cm]{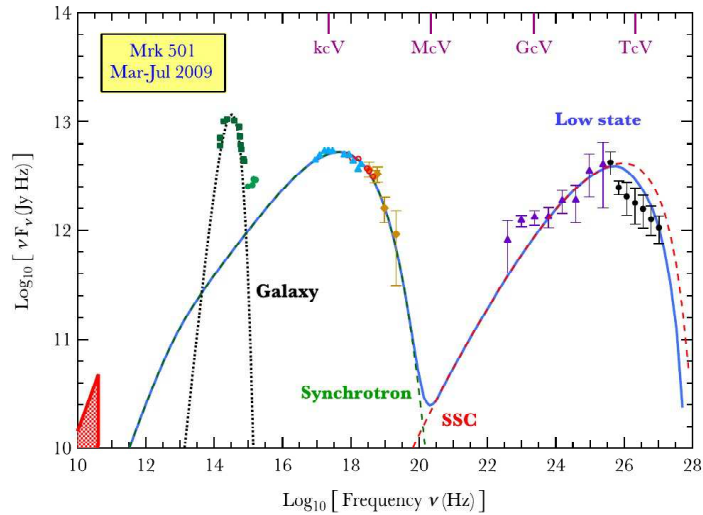}
\caption{Fit to the SED of the HBL Mrk 501 (data from \cite{Abdo11}), with $\eta_1 = 100$ and
$\alpha = 1.5$. From \cite{Baring16}. }
\label{Mrk501}
\end{figure}   

Figure \ref{Mrk501} shows a fit to the SED of the high-frequency peaked BL Lac object (HBL) Mrk 501
\citep[data from][]{Abdo11}, which exhibits its characteristic synchrotron peak at soft X-ray energies.
This, along with efficient acceleration of particles out of the thermal pool, could be achieved with
a choice of $\eta_1 = 100$ and $\alpha = 1.5$, i.e., $\eta (\gamma) = 100 \, \gamma^{0.5}$. Other parameters
are typical of leptonic synchrotron self-Compton fits to HBLs in general and Mrk 501 in particular. For
details see \cite{Baring16}. 

\begin{figure}[H]
\centering
\includegraphics[width=10cm]{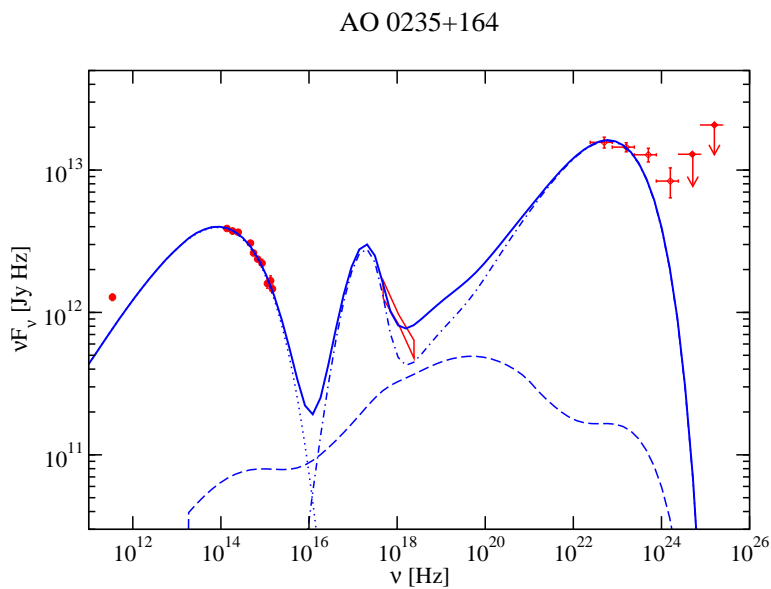}
\caption{Fit to the SED of the LBL AO 0235+164 (data from \cite{Ackermann12}), with $\eta_1 = 225$ and
$\alpha = 3$. Note that the soft X-ray excess (Big Blue Bump) is reproduced by the bulk Compton emission
of the thermal electron population. From \cite{Baring16}. }
\label{AO0235}
\end{figure}   

An example of a low-frequency-peaked BL Lac object (LBL) is shown in Figure \ref{AO0235} \citep[with data
from][]{Ackermann12}. In this case, a much more extreme energy dependence of $\eta$ is required in 
order to reproduce the observed synchrotron peak in the IR: $\eta_1 = 225$ and $\alpha = 3$, i.e., 
$\eta (\gamma) = 225 \, \gamma^2$. In this case, the non-relativistic limit $\eta_1$ is very well 
constrained by the proposition that the soft X-ray excess indicated by the steep {\it Swift}-XRT
2 -- 10~keV spectrum, is produced by the bulk Compton mechanism \citep{Sikora97}, where the thermal
population of electrons Compton up-scatters the external radiation field (dominated by Broad Line
Region emission) by a factor $\sim \Gamma^2$, where $\Gamma$ is the bulk Lorentz factor of the 
post-shock emission region. 

\begin{figure}[H]
\centering
\includegraphics[width=10cm]{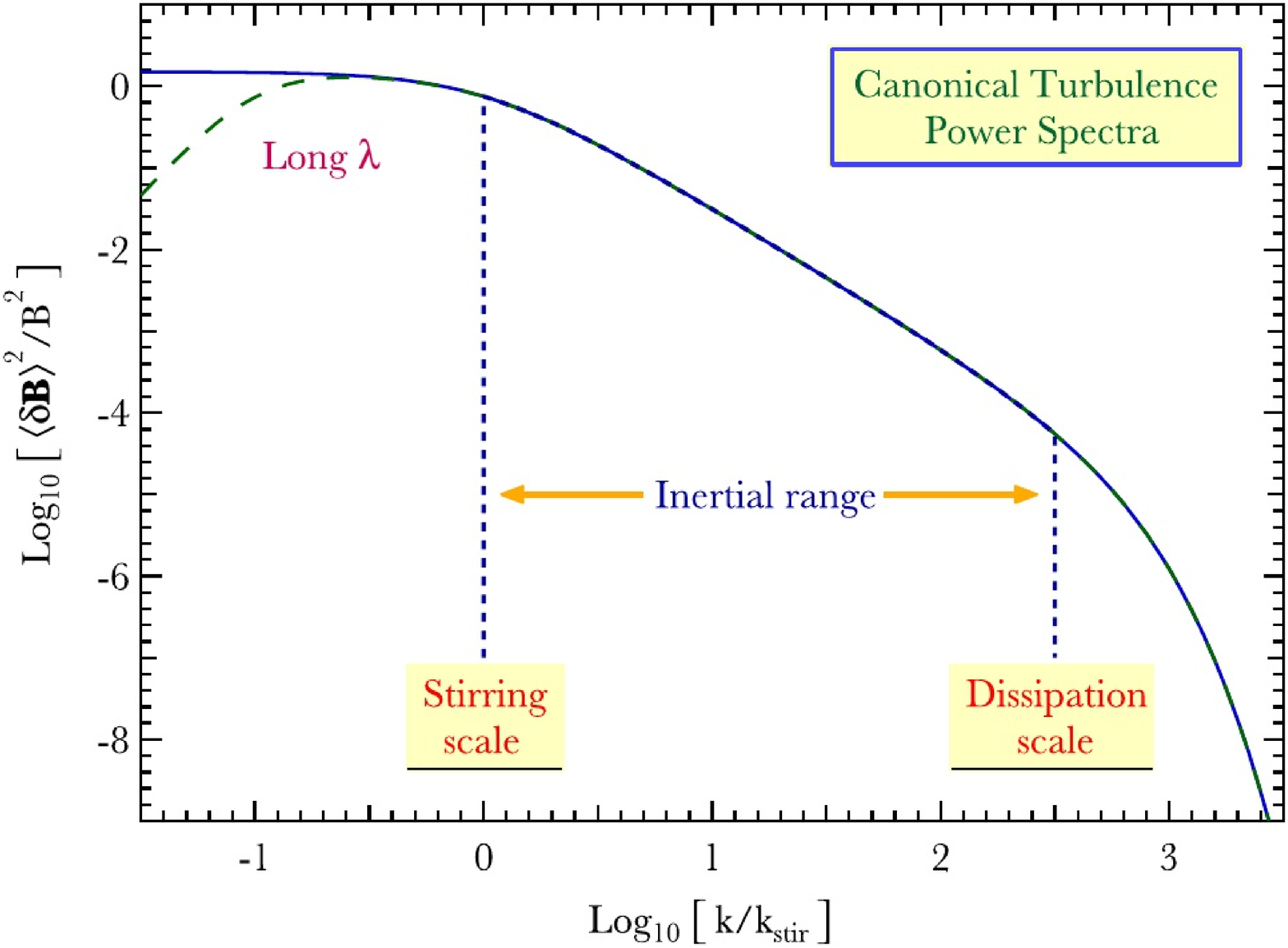}
\caption{Sketch of a magneto-hydrodynamic turbulence spectrum plausibly expected in the post-shock
region. Waves are excited at the longest wavelengths corresponding to the stirring scale, and
cascade down to smaller scales, down to the dissipation scale. From \cite{Baring16}. }
\label{Turbulence}
\end{figure}   

The implied strong energy dependence of the pitch-angle mean free path allows us to diagnose 
the characteristics of hydromagnetic turbulence in the blazar jet. Specifically, pitch-angle
scattering is most efficient for particles interacting gyro-resonantly with plasma waves of
the same (co-moving) frequency as their own gyro-frequency. This translates into gyro-resonant
particle-wave interactions with waves having a wave number $k_{\parallel}$ in the propagation
direction along the guiding magnetic field, given by $k_{\parallel} = \omega_B / (\gamma \, 
v \, \vert\mu\vert)$ where $\omega_B$ is the particles gyro-frequency, $v$ its velocity, and
$\mu$ the cosine of its pitch angle. To first order, this implies that the resonant wave length
for a particle with energy $\gamma$ scales as $\lambda_{\rm res} \propto \gamma$. Consequently,
the highest-energy particles will interact primarily with the longest wavelength plasma waves
(smallest $k_{\parallel}$). Thus, two effects are expected to combine to suppress such gyro-resonant
interactions for very long wavelengths: (1) For wavelengths approaching (and exceeding) the 
``stirring scale'' on which plasma turbulence is induced, the energy density contained in such
waves is rapidly decreasing with increasing wavelengths (see Figure \ref{Turbulence}); 
(2) the highest energy particles will probe large volumes behind the shock front, where 
the level of turbulence is expected to decrease with distance from the shock. These 
effects provide a natural explanation of the strong energy dependence of the pitch-angle 
mean free path which we find in our analyses. The internal consistency of this picture
provides support for DSA as the dominant particle acceleration mechanism in blazar jets.

\section{\label{PAswings}Polarization-Angle Swings}

As an alternative to the purely geometry-based models of helical pattern motions for the
optical PA swings observed in several blazars, as discussed in Section \ref{intro}, correlated 
with multi-wavelength flaring activity, Zhang et al. \cite{Zhang14,Zhang15} have proposed a
model which does not require any asymmetric jet features and/or pattern motions. It is
the first model that consistently considers particle dynamics and time- and polarization-dependent
radiation transport, along with all relevant light-travel time effects. Most importantly, 
Zhang et al. \cite{Zhang14} have shown that careful consideration of all light travel time
and radiation-transfer effects naturally leads to $\sim 180^o$ PA rotations when a shock 
propagates through an active region in a jet pervaded by a helical magnetic field. 

\begin{figure}[H]
\centering
\includegraphics[width=11cm]{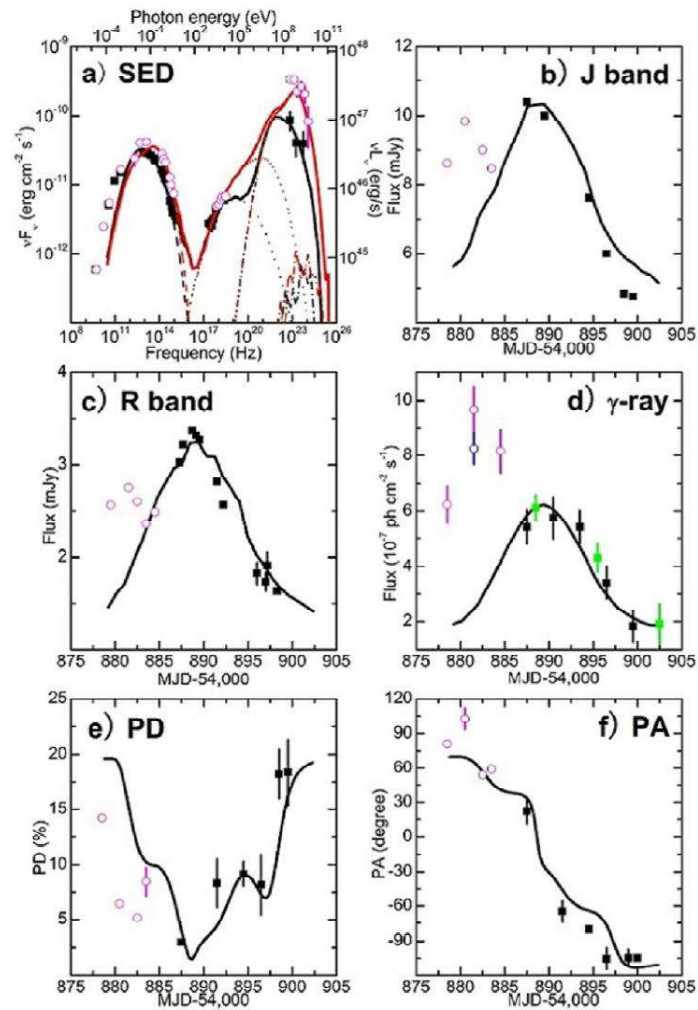}
\caption{The multiwavelength flare + PA swing event of 3C279 in 2009 \cite{Abdo10}: 
Simultaneous fits to snap-shot SEDs (a), multi-wavelength light curves (b -- d), and
time-dependent optical polarization characteristics (e -- f). From \cite{Zhang15}. }
\label{3C279}
\end{figure}   

A spectacular success of this model is provided by its application to the prominent PA rotation
+ multi-wavelength flare event of 3C279 in 2009 \citep{Abdo10}, as presented in \cite{Zhang15}. 
Here, the authors have presented, for the first time, fits to snap-shot SEDs, multi-wavelength
light curves, and time-dependent optical synchrotron polarization signatures (Polarization Degree
and Polarization Angle), within one consistent model, properly treating particle dynamics and
polarization-dependent radiation transfer. A key result of this study was that these fits 
required a decreasing magnetic field strength along with enhanced particle acceleration,
indicative of magnetic energy dissipation as the driving mechanism for the observed flaring
activity. For details of the simulations and the parameters required for the fit presented in
Figure \ref{3C279}, see \cite{Zhang15}.

\section{\label{leptohad}A Polarization Dependent Multi-Zone Lepto-Hadronic Internal Shock Model}

While the electron-synchrotron origin of the low-energy (radio -- optical/UV/X-ray) SED component 
of blazars is well established, the nature of the high-energy (X-ray -- $\gamma$-ray) emission is
less clear. Both leptonic scenarios, where the $\gamma$-rays result from Compton scattering of
various radiation fields by the same leptons that produce the low-energy synchrotron emission,
and hadronic scenarios, where the high-energy emission is produced by ultrarelativistic protons
(via proton synchrotron radiation and/or photo-pion production and subsequent pion decay), are
viable and being actively investigated. However, due to the significantly more complex nature
of hadronic interactions (and subsequent pion and muon decay and cascading), hadronic models
are only now beginning to be investigated in a time-dependent manner, and to the author's
knowledge, all published works on hadronic blazar models are single-zone in nature, most of
them time-independent \citep[e.g.][]{MB92,MP01,Muecke03,PM12,Boettcher13,Mastichiadis13,Cerruti15,Petropoulou15}. 

A fully time-dependent single-zone lepto-hadronic blazar model had been developed recently by 
Diltz et al. \cite{Diltz15}, where the particle dynamics of all decay products is properly taken
into account, including pion and muon synchrotron radiation. The proton evolution and proton
synchrotron radiation routines of that model have recently been incorporated into the polarization-dependent
radiation-transfer model described in the previous section. This provides us with the first 3D
time- and polarization-dependent shock-in-jet model incorporating proton synchrotron radiation.
As detailed cross-zone radiation transport is not included in the model, synchrotron self-Compton
radiation as well as photo-pion induced processes are neglected, which restricts the model to a
parameter regime in which the magnetic energy density greatly exceeds the co-moving photon
energy density (for detailed parameter estimates resulting from this restriction, see \cite{Zhang16}). 

\begin{figure}[H]
\centering
\includegraphics[width=12cm]{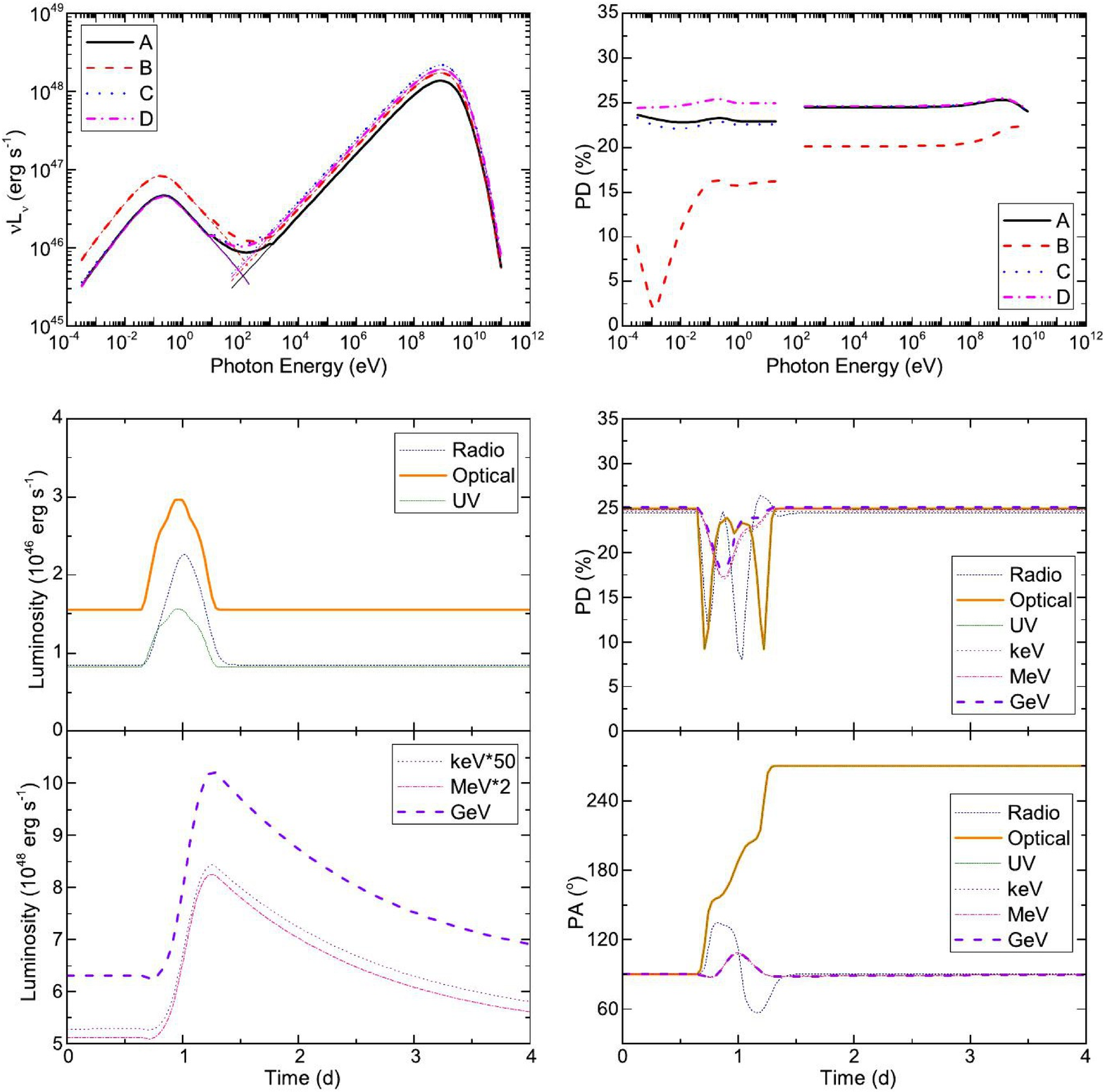}
\caption{Radiation and polarization signatures from a proton-synchrotron dominated lepto-hadronic
shock-in-jet model, in a scenario where flaring activity is initiated by more efficient particle
acceleration, accompanied by a decrease of the magnetic field (magnetic energy dissipation) and
where the light crossing time across the jet is shorter than the characteristic radiative cooling
time of the highest-energy protons. Top left: Snap-shot SEDs for (A) the pre-flare equilibrium
state, (B) a period shortly before the peak of the flare, (C) the peak of the flare, and (D) during
the decay phase after the flare peak; top right: Polarization Degree (PD) as a function of photon 
energy for the same epochs as the SEDs shown in the top-left panel; center and bottom left: 
Multi-wavelength light curves; center and bottom right: Time-dependent Polarization Degree (PD)
and Polarization Angle (PA) at various photon energies. From \cite{Zhang16}. }
\label{3Dhad}
\end{figure}

In \cite{Zhang16}, we are presenting a detailed parameter study of our proton-synchrotron
shock-in-jet model. We find that the variability characteristics of the polarization signatures
(Polarization Degree and Polarization Angle) depend critically on the relation between the
light crossing time across the active region, $t_{\rm cr}$, and the radiative cooling time 
scale $t_{\rm cool}$ of the highest-energy protons. Figure \ref{3Dhad} illustrates the results 
in the case $t_{\rm cr} < t_{\rm cool}$ for a flaring scenario in which more efficient particle 
acceleration is accompanied by a reduction of the magnetic field (magnetic energy dissipation). 
The jet is pervaded by a helical magnetic field with a similar baseline configuration as used
in \cite{Zhang14,Zhang15} to reproduce PA swings correlated with multi-wavelength flaring activity
in a leptonic model context.

The baseline parameter set leads to an SED that is characteristic of a low-frequency peaked
blazar so that, in this model, the X-ray emission is dominated by the same proton synchrotron
radiation component responsible for the $\gamma$-rays. The figure illustrates that this scenario 
leads to significant multi-wavelength flaring activities and that the high-energy (X-ray through
$\gamma$-ray) emission is expected to be highly polarized, to a similar degree as the optical 
(electron-synchrotron) emission during the quiescent state. 

Most importantly, the bottom-left panel of Figure \ref{3Dhad} shows that the PA of the high-energy
emission is expected to stay relatively stable throughout the shock-induced flare, while the optical
emission exhibits the well-known 180$^o$ PA swing. The reason for this is the significantly longer
radiative cooling time scale for protons, compared to electrons, due to which a much larger region 
behind the shock remains active. This is also reflected in the significantly longer decay time scale
of the X-ray through $\gamma$-ray emission compared to the electron-synchrotron (radio --- UV) emission.

This has important consequences for the prospects of future X-ray or $\gamma$-ray polarimetry
missions (e.g., XIPE, PolSTAR, PANGU) to detect high-energy polarization from blazars. During
the required long integration time of these instruments to potentially detect X-ray (or $\gamma$-ray)
polarization from blazars, the optical PA may very well vary by significant amounts and possibly exhibit
large-angle PA swings. If the X-ray / $\gamma$-ray PA were behaving in the same way, and this was
not properly accounted for in the data analysis of high-energy polarimeters, this would completely 
smear out any polarization signatures. Instead, we here find that quite plausibly, the PA of
proton-synchrotron dominated X-ray and $\gamma$-ray emission may remain stable even if the optical
PA exhibits large-angle swings, thus greatly enhancing the chances of a positive detection of these
high-energy polarization signatures. As pointed out by Zhang \& B\"ottcher \cite{ZB13}, the possibly
substantial high-energy polarization degrees predicted by hadronic models are in stark contrast to 
the very low degree of X-ray and $\gamma$-ray polarization expected in leptonic models. Thus, our
results re-inforce future prospects of using X-ray and $\gamma$-ray polarization as a diagnostic 
to distinguish leptonic from hadronic high-energy emission in blazars.

\acknowledgments{M.B. acknowledges support through the South African Research Chairs Initiative of 
the South African Department of Science and Technology and the National Research Foundation\footnote{Any 
opinion, finding and conclusion or recommendation expressed in this material is that of the authors, 
and the NRF does not accept any liability in this regard.} under SARChI grant no. 64789. This work
has also been supported by NASA through Astrophysics Theory Program grant NNX10AC79G.}

\bibliographystyle{mdpi}

\renewcommand\bibname{References}


\begin{thebibliography}{999}

\bibitem{Abdo10}
Abdo, A. A., et al., A change in the optical polarization associated with a $\gamma$-ray flare in the blazar
3C 279. {\em Nature} {\bf 2010}, {\em 463}, 919-923

\bibitem{Abdo11}
Abdo, A. A., et al., Inights into the high-energy $\gamma$-ray emission of Markarian 501 from extensive 
multifrequency observations in the Fermi era. {\em ApJ} {\bf 2011}, {\em 727}, 129.

\bibitem{Ackermann12}
Ackermann, M., et al., Multiwavelength observations of blazar AO 0235+164 in the 2008 - 2009 flaring state.
{\em ApJ} {\bf 2012}, {\em 751}, 159.

\bibitem{Baring16}
Baring, M. G., B\"ottcher, M., \& Summerlin, E. J., Probing acceleration and turbulence at relativistic
shocks in blazar jets. {\em MNRAS} {\bf 2016}, submitted.

\bibitem{Blinov16}
Blinov, D., et al., RoboPol: optical polarization-plane rotations and flaring activity in blazars. 
{\em MNRAS} {\bf 2016}, {\em 457}, 2252-2262.

\bibitem{BD10}
B\"ottcher, M., \& Dermer, C. D., Timing signatures of the internal-shock model for blazars. {\em ApJ}
{\bf 2010}, {\em 711}, 445-460.

\bibitem{Boettcher13}
B\"ottcher, M., Reimer, A., Sweeney, K., \& Prakash, A., Leptonic and hadronic modeling of Fermi-detected
blazars. {\em ApJ} {\bf 2013}, {\em 768}, 54.

\bibitem{Cerruti15}
Cerruti, M., Zech, A., Boisson, C., \& Inoue, S., A hadronic origin for ultra-high-frequency-peaked BL Lac objects.
{\em MNRAS} {\bf 2015}, {\em 448}, 910-927.

\bibitem{Chen11}
Chen, X., Fossati, G., Liang, E. P., \& B\"ottcher, M., Time-dependent simulations of multiwavelength
variability of the blazar Mrk 421 with a Monte Carlo multizone code. {\em MNRAS} {\bf 2011}, {\em 416},
2368-2387.

\bibitem{Chen12}
Chen, X., Fossagi, G., B\"ottcher, M., \& Liang, E. P., Time-dependent simulations of emission from the FSRQ
PKS 1510-089: multiwavelength variability of external Compton and synchrotron self-Compton models. {\em MNRAS}
{\bf 2012}, {\em 424}, 789-799.

\bibitem{Diltz15}
Diltz, C. S., B\"ottcher, M., \& Fossati, G., Time dependent hadronic modeling of flat spectrum radio quasars.
{\em ApJ} {\bf 2015}, {\em 802}, 133.

\bibitem{Ghisellini10}
Ghisellini, G., et al., General physical properties of bright Fermi blazars. {\em MNRAS} {\bf 2010}, {\em 402},
497-518.

\bibitem{Graff08}
Graff, P. B., Georganopoulos, M., Perlman, E. S., \& Kazanas, D., A multizone model for simulating the
high-energy variability of TeV blazars. {\em ApJ} {\bf 2008}, {\em 689}, 68-78.

\bibitem{JB11}
Joshi, M., \& B\"ottcher, M., Time-dependent radiation transfer in the internal shock model scneario for
blazar jets. {\em ApJ} {\bf 2011}, {\em 727}, 21

\bibitem{Larionov13}
Larionov, V. M., et al., The outburst of the blazar S5 0716+71 in 2011 October: shock in a helical jet. {\em ApJ}
{\bf 2013}, {\em 768}, 40.

\bibitem{MB92}
Mannheim, K., \& Biermann, P. L., Gamma-ray flaring of 3C 279 - a proton-initiated cascade in the jet?
{\em A\&A} {\bf 1992}, {\em 253}, L21-L24.

\bibitem{MG85}
Marscher, A. M., \& Gear, W. K., Models for high-frequency radio outbursts in extragalactic sources, with
application to the early 1983 millimeter-to-infrared flare of 3C 273. {\em ApJ} {\bf 1985}, {\em 298}, 114-127.

\bibitem{Marscher08}
Marscher, A. P., et al. The inner jet of an active galactic nucleus revealed by a radio-to-$\gamma$-ray outburst.
{\em Nature} {\bf 2008}, {\em 452}, 966-969.

\bibitem{Marscher14}
Marscher, A. P., Turbulent, extreme multi-zone model for simulating flux and polarization variability in blazars.
{\em ApJ} {\bf 2014}, {\em 780}, 87.

\bibitem{Mastichiadis13}
Mastichiadis, A., Petropoulou, M., \& Dimitrakoudis, S., Mrk 421 as a case study for TeV and X-ray variability in
leptohadronic models. {\em MNRAS} {\bf 2013}, {\em 434}, 2684-2695.

\bibitem{Mimica04}
Mimica, P., Aloy, M., M\"uller, E., \& Brinkmann, W., Synthetic X-ray light curves of BL Lacs from relativistic
hydrodynamic simulations. {\em A\&A} {\bf 2004}, {\em 418}, 947-958.

\bibitem{MP01}
M\"ucke, A., \& Protheroe, R. J., A proton synchrotron blazar model for flaring in Markarian 501. {\em Astropart. 
Phys.} {\bf 2001}, {\em 15}, 121-136.

\bibitem{Muecke03}
M\"ucke, A., et al., BL Lac objects in the synchrotron proton blazar model. {\em Astropart. Phys.} {\bf 2003},
{\em 18}, 593-613.

\bibitem{PM12}
Petropoulou, M., \& Mastichiadis, A., On proton synchrotron blazar models: the case of quasar 3C279. {\em MNRAS}
{\bf 2012}, {\em 426}, 462-472.

\bibitem{Petropoulou15}
Petropoulou, M., et al., Photohadronic origin of $\gamma$-ray BL Lac emission: implications for IceCube neutrinos.
{\em MNRAS} {\bf 2015}, {\em 448}, 2412-2429.

\bibitem{Sikora97}
Sikora, M., Madejski, G., Moderski, R., \& Poutanen, J., Learning about active galactic nucleus jets from spectral
properties of blazars. {\em ApJ} {\bf 1997}, {\em 484}, 108-117.

\bibitem{Sokolov04}
Sokolov, A., Marscher, A. P., \& McHardy, I. M., Synchrotron self-Compton model for rapid nonthermal flares in
blazars with frequency-dependent time lags. {\em ApJ} {\bf 2004}, {\em 613}, 725-746.

\bibitem{SM05}
Sokolov, A., \& Marscher, A. P., External Compton radiation from rapid nonthermal flares in blazars.
{\em ApJ} {\bf 2005}, {\em 629}, 52-60.

\bibitem{Spada01}
Spada, M., Ghisellini, G., Lazzadi, D., \& Celotti, A., Internal shocks in the jets of radio-loud quasars.
{\em MNRAS} {\bf 2001}, {\em 325}, 1559-1570.

\bibitem{SB12}
Summerlin, E. J., \& Baring, M. G., Diffusive acceleration of particles at oblique, relativistic magnetohydrodynamic
shocks. {\em ApJ} {\bf 2012}, {\em 745}, 63.

\bibitem{ZB13}
Zhang, H., \& B\"ottcher, M., X-ray and gamma-ray polarization in leptonic and hadronic jet models of blazars.
{\em ApJ} {\bf 2013}, {\em 774}, 18.

\bibitem{Zhang14}
Zhang, H., Chen, X., \& B\"ottcher, M., Synchrotron polarization in blazars. {\em ApJ} {\bf 2014}, {\em 789}, 66.

\bibitem{Zhang15}
Zhang, H., et al. Polarization swings reveal magnetic energy dissipation in blazars. {\em ApJ} {\bf 2015},
{\em 804}, 58.

\bibitem{Zhang16}
Zhang, H., Diltz, C. S., \& B\"ottcher, M., Radiation and polarization signatures of a 3C multi-zone, 
time-dependent hadronic blazar model. {\em ApJ} {\bf 2016}, submitted.

\end{thebibliography}


\end{document}